\documentclass[prl,twocolumn,amsmath,amssymb]{revtex4}

\usepackage{graphicx}

\begin{document}

\title{Plasmon-phonon coupling in large-area graphene dot and antidot arrays}

\author{Xiaolong~Zhu$^{1,2}$, Weihua~Wang$^{1,2}$, Wei~Yan$^{1,2}$, Martin~B.~Larsen$^{3,2}$, Peter~B{\o}ggild$^{3,2}$,\\ Thomas~Garm~Pedersen$^{4,2}$, Sanshui~Xiao$^{1,2}$, Jian~Zi$^4$, and N.~Asger~Mortensen$^{1,2}$}

\email{asger@mailaps.org}

\affiliation{\tiny $^1$Department of Photonics Engineering, Technical University of Denmark, DK-2800 Kgs. Lyngby, Denmark\\$^2$Center for Nanostructured Graphene (CNG), Technical University of Denmark, DK-2800 Kgs. Lyngby, Denmark\\$^3$Department of Micro and Nanotechnology, Technical University of Denmark, DK-2800 Kgs. Lyngby, Denmark\\$^4$Department of Physics and Nanotechnology, Aalborg University, DK-9220 Aalborg, Denmark\\$^5$Department of Physics, Key Laboratory of Micro- and Nano-Photonic Structures (Ministry of
Education) and State Key Laboratory of Surface Physics, Fudan University
Shanghai 200433, China}

\begin{abstract}
Nanostructured graphene on SiO$_2$ substrates pave the way for enhanced light-matter interactions and explorations of strong plasmon-phonon hybridization in the mid-infrared regime. Unprecedented large-area graphene nanodot and antidot optical arrays are fabricated by nanosphere lithography, with structural control down to the sub-100 nanometer regime. The interaction between graphene plasmon modes and the substrate phonons is experimentally demonstrated and structural control is used to map out the hybridization of plasmons and phonons, showing coupling energies of the order 20\,meV. Our findings are further supported by theoretical calculations and numerical simulations.
\end{abstract}

\maketitle

Plasmon polaritons, the light-driven collective oscillation of electrons, provide the foundation for various applications ranging from metamaterials, plasmonics, photocatalysis to biological sensing~\cite{Maier:2007}. Owing to the two dimensional (2D) feature of the excitations, plasmon polaritons supported by graphene are of particular interest~\cite{Koppens:2011,Fei:2012,Chen:2012}. Tight mode confinement, long propagation distance, and remarkable electrostatic tunability of graphene plasmon polaritons lead to new applications for waveguides, modulators and super-lenses~\cite{Vakil:2011}~\cite{Zhu:2013b}~\cite{Lu:2013}. Plasmon polaritons in a graphene sheet can be excited by advanced near-field scattering microscopy~\cite{Fei:2012,Chen:2012}, dielectric subwavelength grating coupler~\cite{Zhan:2012,Zhu:2013a}, or nanoscale patterning~\cite{Ju:2011,Yan:2012,Nikitin:2012,Thongrattanasiri:2012,Fang:2013}, while the dispersion of graphene plasmons may be influenced by the interaction of electrons and the surface optical phonons of the polar substrate~\cite{Fei:2011,Daas:2011,Koch:2010}. Using angle-resolved reflection electron-energy-loss spectroscopy, strong plasmon-phonon coupling has been confirmed in epitaxial graphene placed on the silicon carbide substrate~\cite{Liu:2010}. The plasmon-phonon interaction leads to additional damping channel for graphene plasmons by the mid-infrared transmission measurement assisted by an infrared microscope coupled to a Fourier-transform infrared spectrometer~\cite{Yan:2013}. More recently, phonon-induced transparency in bilayer graphene nanoribbon has been demonstrated, resulting in a maximum slow light factor of around 500~\cite{Yan:2013b}.

In this paper, we demonstrate an effective approach for patterning graphene sheets into large-area ordered graphene nanostructures by combining nanosphere lithography (NSL) with O$_2$ reactive ion etching (RIE). Without high-cost and low-throughput lithographic patterning and sophisticated instruments, we realize nanoscale graphene dot and antidot arrays with dimensions down to 100\,nm via a single-step fabrication process. Mid-infrared (5 -– 15\,$\rm\mu$m) plasmons in graphene dot and antidot arrays can be controlled through either the structure size or plasmon-phonon hybridization (when using a polar substrate). Measured reflection spectra illustrate the excitation of plasmons in the graphene nanostructures and their coupling with surface polar phonons in the SiO$_2$ substrate. These results are further supported by theoretical predictions and numerical simulations. Our study has enabled large-area fabrication of graphene nanostructures and the experimental investigations of graphene plasmon-phonon coupling are especially important for graphene-based electro-optical devices.

\begin{figure}[b!]
\begin{center}
\includegraphics[width=1\columnwidth]{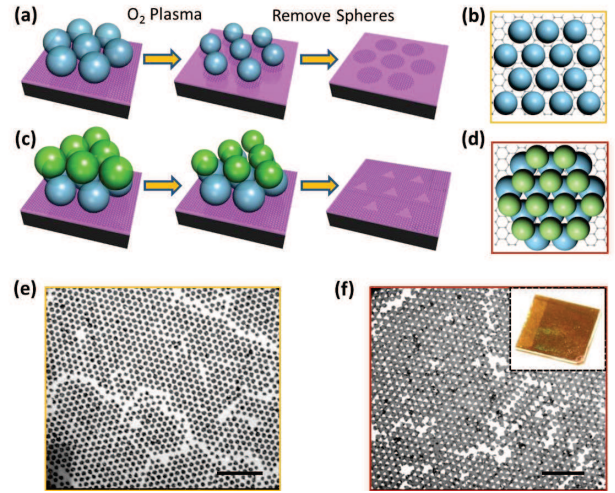}
\end{center}
\caption{Fabrication processes of graphene nanodot and antidot arrays. ({\bf a}) and ({\bf b}) The top-view schematics of the nanosphere templates for graphene nanodot. ({\bf c}) and ({\bf d}) The top-view schematics of the nanosphere templates for graphene antidot. ({\bf e}) and ({\bf f}) SEM images for the corresponding large-area graphene nanodot and antidot arrays. Scale bars: 10\,$\rm\mu$m. The inset in ({\bf f}) is a photograph of a graphene antidot array with a 1\,cm$^2$ area.\label{fig1}}
\end{figure}

\begin{figure}[t!]
\begin{center}
\includegraphics[width=1\columnwidth]{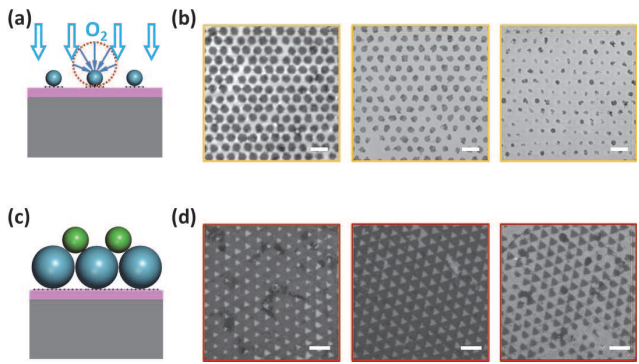}
\end{center}
\caption{Side-view schematic of the template for fabricating graphene dots and antitods. ({\bf a}) Side-view schematic of the template for fabricating graphene dots, illustrating the
shrinking process of the spheres under O$_2$ plasma explosion. ({\bf b}) SEM images of graphene dot
arrays under 10\,s (left), 20\,s (middle) and 30\,s (right) plasma etching, respectively, by using
1500\,nm (diameter) sphere array as the template. ({\bf c}) Side-view schematic of the template for
fabricating graphene antidots. ({\bf d}) SEM image of graphene triangular antidots and dots
obtained under 20\,s (left), 30\,s (middle) and 40\,s (right) plasma etching. The scale bars are 2\,$\rm\mu$m.\label{fig2}}
\end{figure}

We used a self-assembled nanosphere template and RIE processes to fabricate both graphene dot and antidot arrays. The graphene grown by chemical vapor deposition (CVD) on copper was firstly transferred to SiO$_2$/Si substrates by a wet transfer method~\cite{Li:2009}. Then, a monolayer of polystyrene (PS) spheres was prepared on the graphene-coated substrates using the capillary effect~\cite{Sun:2010}, as illustrated in Figure 1a. While the contact with polymer residues for instance originating from photoresist is known to compromise the electrical properties of graphene~\cite{Zhou:2010}, the point-like contact to the spherical template is expected to leave the graphene in a relatively clean state. Then, O$_2$ plasma was applied to transform the closely packed PS nanosphere monolayer into arrays of separated nanospheres. Areas of the graphene that are not shadowed by the nanospheres will be etched away (Figure 1b). By sonication in ethanol for a few minutes, the PS spheres can be easily removed. Eventually, periodically arranged graphene dots are realized, as shown in a scanning electron microscopy (SEM) micrograph in Figure 1e. More importantly, by carefully controlling the concentration of the nanosphere solution, a bilayer array of close-packed PS spheres could be obtained, as illustrated in Figure 1c. Because of the shadow effect, a graphene antidot array will be realized after the etching process (see Figure 1d). After exposure to O$_2$ plasma for a few seconds, hexagonally arranged triangular holes with sharp corners, almost identical in size and shape, are obtained in large areas with excellent periodicity and reproducibility, as shown in the SEM micrograph (Figure 1f). It should be noted that the NSL applied here is well-suited for fabricating large-scale inexpensive structures: by scaling the nanosphere diameters, the dimensions of the structures can be further tuned. The antidot array has been suggested as a means of opening a tunable electronic bandgap, controlling the in-plane charge-carrier transport in graphene~\cite{Pedersen:2008,Ouyang:2011,Zhu:2010}. However, large-scale experimental fabrication is still problematic. Antidot arrays have been fabricated by inherently slow serial fabrication methods such as electron beam lithography~\cite{Shi:2011}, or by block-copolymer lithography, while they are challenging in terms of reproducibility and complexity due to stringent requirements to surface properties, polymer thin film homogeneity and processing conditions~\cite{Segalman:2005}. Here, we provide a fast, simple and robust route towards large-area graphene antidot arrays for optoelectronics.

\begin{figure}[b!]
\begin{center}
\includegraphics[width=1\columnwidth]{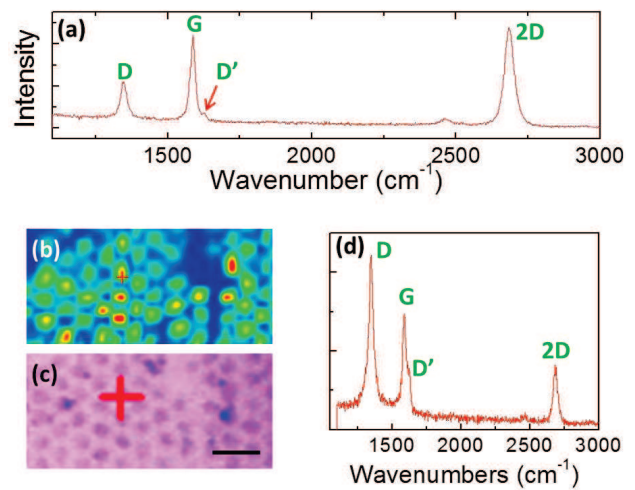}
\end{center}
\caption{Raman spectra and Raman-imaging of samples. ({\bf a}) Raman spectra of as-transferred CVD-graphene on SiO$_2$/Si substrate. ({\bf b}) Raman image of graphene nanodots plotted by extracting the integrated intensity of 2D band. ({\bf c}) \emph{In-situ} optical snapshot on the same part of the sample as in ({\bf b}). The scale bar is 2\,$\rm\mu$m. ({\bf d}) Micro-Raman spectra of specific graphene dots indicated by the red cross in ({\bf b}) and ({\bf c}).\label{fig3}}
\end{figure}

With a single layer NSL template, the shape of the graphene nanostructures can still be varied and continually tuned via the etching process. By carefully controlling the O$_2$ plasma etching time, the close-packed spheres will become separated while the size of PS spheres will be gradually reduced. The graphene sheet within the gaps between spheres was selectively etched away together with the shrinking part of spheres by exposure to the O$_2$ plasma, as illustrated in Figure 2a. Highly ordered arrays of graphene nanodots with tunable diameters down to 100\,nm were in this way fabricated. Figure 2b shows the SEM pictures of the graphene dot arrays resulting from 10\,s (left), 20\,s (middle), and 30\,s (right) plasma etching, respectively, by using an array of 1500\,nm diameter spheres as the template. For the case of the bilayer-sphere template, the top layer is etched faster than the lower one because of the relative shadowing, as illustrated in Figure 2c. We produced graphene antidots with an equilateral triangular shape after the anisotropic RIE of the double layer of 1500\,nm PS spheres, which were arranged into a hexagonal array. Figure 2d shows SEM micrographs of the graphene antidot arrays fabricated under different plasma etching time. When RIE was further performed, the graphene antidots will be enlarged and finally become separated triangular dots.

\begin{figure}[t!]
\begin{center}
\includegraphics[width=1\columnwidth]{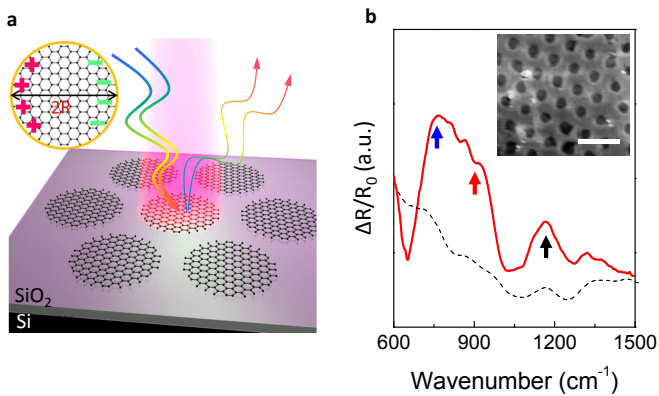}
\end{center}
\caption{Graphene nanodot spectra. ({\bf a}) SEM image of a typical array of graphene dots. The radius of the dots is $R\sim$100\,nm. The scale bar is 1\,$\rm\mu$m. ({\bf b}) Relative reflection spectra of the corresponding nanodot array (solid curve) and the unstructured graphene (the dashed curve) on a SiO$_2$/Si substrate. The inset is a schematic illustration of dipole oscillation in a graphene dot.\label{fig4}}
\end{figure}

Raman spectroscopy is a powerful non-destructive technique to explore the properties and structure of graphene samples, such as the number of layers, atomic structure, doping level, strain and disorder~\cite{Das:2008,Ni:2008,Ferrari:2006}. Here, we employ micro-Raman spectroscopy to characterize the quality of the samples. The Raman spectrum excited at 532\,nm of the as-transferred CVD-graphene is shown in Figure 3a. Prominent features with Raman shifts of the G peak around 1585\,cm$^{-1}$ and the 2D peak around 2683\,cm$^{-1}$ are observed. Furthermore, the 2D peak exhibits a single Lorentzian shape, which is the signature of a monolayer graphene. The low-intensity peak around 1345\,cm$^{-1}$ is attributed to the D band of carbon, which is activated by the presence of defects in the transferred CVD graphene, such as vacancies or grain boundaries, as well as adsorbents remaining from the graphene transfer process. The shoulder peak locating at 1625\,cm$^{-1}$, so-called the D' band, is also related to defects~\cite{Eckmann2013}. The effect of doping on the Raman frequency and full-width at half-maximum (FWHM) of the G-band, as well as the intensity ratio of the G and 2D peaks ($I_{\scriptscriptstyle 2D}/I_{\scriptscriptstyle G}$) has been intensively investigated~\cite{Das:2008}. The large FWHM (G) and small $I_{\scriptscriptstyle 2D}/I_{\scriptscriptstyle G}$ in Figure 3a indicate the high doping level of the CVD-grown graphene.

\begin{figure}[b!]
\begin{center}
\includegraphics[width=1\columnwidth]{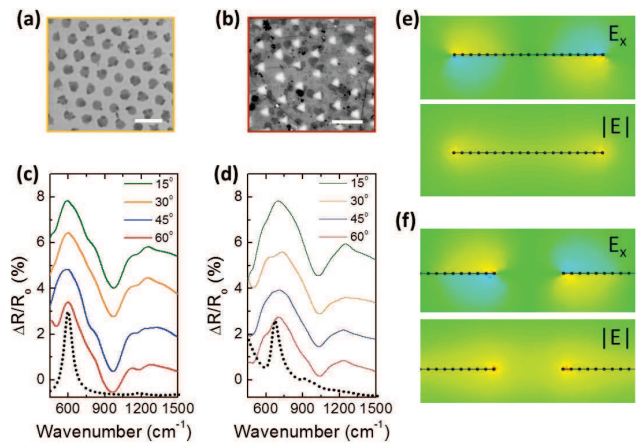}
\end{center}
\caption{Angle-resolved nanodot and antidot spectra. SEM images and the corresponding angle-resolved spectra of graphene dots ({\bf a},{\bf c})
and antidots ({\bf b},{\bf d}) array fabricated by applying single- and double-layer PS spheres as
templates and both exposed to 20\,s of O$_2$ plasma etching. Scale bars in ({\bf a}) and ({\bf b}) are 2\,$\rm\mu$m. The dotted curves in ({\bf c}) and ({\bf d}) are the simulated spectra for the corresponding graphene dot and antidot array by using a constant value as the dielectric function of SiO$_2$ substrate. The electric field distributions at the resonance frequencies of graphene ({\bf e}) dot and ({\bf f}) antidot arrays are also plotted along the high-symmetry $x$-direction through the center of the dot/antidot.\label{fig5}}
\end{figure}

Figure 3b shows a representative micro Raman image of the graphene nanodisk arrays (shown in Figure 1e) obtained by extracting the integrated intensity of the 2D band. The intensity plot reveals a periodic pattern, showing a good agreement with the in-situ optical snapshot in Figure 3c and its corresponding SEM micrograph. From the Raman spectrum of a specific graphene dot labeled by a red cross in the Raman map as well as the microscopic image, we find a clear difference when comparing with the Raman spectrum of the original CVD-graphene. The corresponding D and D' bands are strengthened, which is associated with the
rich variation in the boundaries of graphene disks (zigzag versus armchair termination) and defects introduced by O$_2$ plasma etching (Figure 3d). More importantly, further FWHM decreasing (G) and $I_{\scriptscriptstyle 2D}/I_{\scriptscriptstyle G}$ ratio reduction occur, which may be explained by a promoted charge doping considering oxygen atoms or ions binding with the carbon dangling bonds during the
O$_2$ plasma process. Based on this observation, the Fermi level is possibly decreased further to near -0.45\,eV~\cite{Das:2008}, which is taken into consideration in our numerical simulations in the
following section.

Electromagnetic radiation can couple to plasmon excitations in engineered graphene structures with dimensions much smaller than free-space optical wavelengths~\cite{Thongrattanasiri:2012}. With the aid of the NSL method, we achieve a graphene dot array with disk radius $R\sim$100\,nm, see Figure 4a, when using PS spheres with a diameter down to 600\,nm. The mid-infrared reflection response of the graphene dot array is measured by a set-up consisting of an infrared microscope coupled to a Fourier-transform infrared spectrometer (FTIR). The electromagnetic response of the graphene array is quantified by the relative reflection spectrum $\Delta R/R_0$, where $\Delta R$ is the reflection difference between the light reflected from the graphene arrays and from the bare substrate, and $R_0$ is the reflection from the bare substrate. The $\Delta R/R_0$ spectra for the $R\sim$100\,nm dot array and for the unstructured graphene sample are shown in Figure 4b. The spectrum of the unstructured graphene sheet exhibits small resonance peaks while translational invariance makes the plasmon non-radiative. The oscillation in the dashed line only comes from the intrinsic substrate phonons. For the graphene dot array, the spectral feature evolves into multiple strong resonances and is in sharp contrast to the results for \emph{far-infrared} spectra of graphene microstructures in Ref.\onlinecite{Ju:2011}, whose resonances display a single Lorentzian-shape.

We also characterize the angle-resolved response in the mid-infrared frequency range of graphene dot and antidot structures by FTIR. The dot and antidot structures were fabricated by applying single- and double-layer PS spheres as the templates and both exposed to 20\,s O$_2$ plasma etching. Figure 5a and 5b show the SEM images and Figure 5c and 5d present the
corresponding angle-resolved spectra of graphene dot and antidot arrays on a SiO$_2$ substrate, respectively. The spectra all demonstrate several resonances within our measured frequency
range. In particular, the resonance frequencies of the broad spectral peaks (dips) do not change with the increasing angle, ranging from 15$^\circ$ to 60$^\circ$.

In order to support the measured results, we perform full-wave numerical simulations of Maxwell's equations based on a standard finite-integration technique (CST Microwave Studio). The surface conductivity of graphene is considered within the random-phase approximation (RPA)~\cite{Wunsch:2006,Hwang:2007}. For the substrate, we use for a start the common approach where the SiO$_2$ dielectric function is a constant value ($\varepsilon=$2.1, as in Ref.~\onlinecite{Fang:2013}) and independent of wavelength. We carry out electromagnetic simulations for the graphene dot and antidot arrays with dimensions and lattice symmetries resembling the experimental samples, while on the unitcell level we assumed circularly shaped dots and antidots, for simplicity. As shown by the dotted lines in Figure 5c and 5d, the sharp resonances of both cases can be clearly observed and match well with the corresponding main peaks in the experimental measurements. For antidot arrays being ideal negative images of their dot array counterparts, one expects strong correlations between the two types of spectra according to Babinet's principle. This can indeed be seen when comparing panels 5c and 5d, though the quantitative agreement is slightly relaxed due to the formation of triangularly shaped antidots (rather than a circular cross section as for the disk arrays). The electric field distribution maps of the corresponding resonances of graphene dot and antidot arrays are also calculated and shown in Figure 5e and 5f, respectively. Our simulations reveal that the broad resonance is dominated by an edge located, in-plane dipole moment driven by the horizontal electric field $E_x$. Comparing the simulated and experimental spectra, the latter clearly reveals an extra resonance band, locating at around 1200\,cm$^{-1}$ for both the graphene dot and antidot arrays. Another difference is that the main resonance centered at $\sim$600\,cm$^{-1}$ (for the graphene dot arrays) becomes broader than that anticipated from the simulations. It could be attributed to the band being composed of two underlying peaks, which are located at 600 and 800\,cm$^{-1}$ (as a shoulder peak), respectively.

For graphene nanostructures resting on nonpolar substrates, the plasmon excitations satisfy the quasi-static scaling law and the plasmon resonance frequency is expected to follow a simple $\sqrt{q}$ dispersion in the long wavelength limit predicted by RPA theory for the linear response of a two-dimensional electron gas, including graphene, where $q$ is the plasmon wave vector~\cite{Allen:1977,Batke:1986}. For localized dipole plasmon resonances in graphene dots, the effective wave vector $q_{\rm\scriptscriptstyle dot}$ for the dipole mode is given by $q_{\rm\scriptscriptstyle dot} =1/R$, where $R$ is the radius of the nanodots~\cite{Wang:2012}. Thus, we can simply predict the simulated plasmon frequencies, regardless of the spacing between the dots or the period of the array. When homogeneous graphene is periodically modulated by antidots, the plasmon modes can also be excited once the phase-matching condition is fulfilled, i.e. $q_{\rm\scriptscriptstyle antidot}= K_\parallel + G$, where $K_\parallel$ is the in-plane wave vector of the incident wave and $G$ is a reciprocal lattice vector of the graphene antidot array. It is wellknown that the plasmon modes in graphene nanostructures are strongly confined to the graphene plane. In real-space images of the plasmon field (see Figure 5e and 5f), the plasmon wavelength is very short - more than 40 times smaller than the wavelength of illumination~\cite{Chen:2012}, which implies a very large plasmon wave vector, i.e. $q_{\rm\scriptscriptstyle antidot}\gg K_\parallel$ and consequently $q_{\rm\scriptscriptstyle antidot} \sim G$. Thus, the $q_{\rm\scriptscriptstyle antidot}$ is mainly determined by the lattice period $P$, and can be simply given by $q_{\rm\scriptscriptstyle antidot} \simeq 4\pi/\sqrt{3}P$ for a triangular lattice, where in our case $P=$1500\,nm. One can conclude that the deep subwavelength nature of these graphene plasmons can cause angle-independent resonance bands, which are almost unaffected when varying the relatively small $K_\parallel$. These angle-independent and 'localized' modes in semi-continuous and conductive graphene antidot arrays are quite special compared to plasmonic structures of continuous metals~\cite{Zhu:2013c}.

By using different NSL templates and carefully controlling the etching time~\cite{Cong:2009}, we fabricated graphene nanodots with different radii. Figure 6a presents the relative reflection spectra for graphene nanodots on a SiO$_2$ substrate with $R$ ranging from 80 to 290\,nm, demonstrating multiple resonance peaks and even a weak higher-order resonance within our measured frequency range. In order to ensure the uniformity of the nanostructures and strengthen the resonances~\cite{Yan:2012}, we used 600\,nm (diameter) PS spheres as the template to fabricate the nanodots with $R=$80--160\,nm. All the spectra are characterized at 15$^\circ$ by a FTIR with a reflection unit. Although the lattice disorder caused by the self-assembly method may lead to inhomogeneous broadening and reduce the quality factor of the resonances, we can still identify key features in the spectra. The first feature is that all resonance bands (marked by arrows) are blueshifted as $R$ decreases, but with different rates. The second is the competition in spectral weight, which is transferred from the first peak to the second peak with decreasing $R$. The final feature relates to the line shape: the resonance bands are asymmetric and carry the typical fingerprint of a Fano resonance.

Surface optical (SO) phonons are well known in polar surfaces, acting as a polariton of a bosonic quasi-particle comprised of a photon coupled with a transverse electromagnetic field~\cite{Matz:1981,Korobkin:2006}. When graphene is attached to a polar substrate such as SiO$_2$, the coupling between graphene plasmons and the substrate phonons will lead to plasmon-phonon mode hybridization near their crossing energies~\cite{Fei:2011}, thus multiple resonances are expected~\cite{Fei:2011,Fratini:2008,Hwang:2010}. The complicated hybrid plasmon-phonon dispersions in graphene on SiO$_2$ can be theoretically predicted. Within the theoretical framework, this calls for relaxing our initial assumption of constant substrate dielectric function. This assumption is indeed inaccurate for polar SiO$_2$, particularly in the mid-infrared range, where there are two resonances in the dielectric function (corresponding to two phonon modes), leading to a negative dielectric function at specific frequencies. Our extended calculations and simulations include interactions with the relevant substrate surface optical phonons of SiO$_2$ at around 800 and 1100\,cm$^{-1}$ by considering two \emph{reststrahlen} bands in the dielectric function.

The dispersion of the coupled plasmon-phonon surface modes and the anticrossing energy splitting (see also Supplementary Information) can be defined as~\cite{Fratini:2008}
\begin{equation}
\frac{\varepsilon_1(\omega)}{\sqrt{q^2- \varepsilon_1(\omega)\omega^2/c^2}}+\frac{\varepsilon_2(\omega)}{\sqrt{q^2- \varepsilon_2(\omega)\omega^2/c^2}}=-\frac{i\sigma(q,\omega)}{\omega\varepsilon_0}
\end{equation}
which includes the conductance of the graphene layer, $\sigma(q,\omega)$, as well as the dielectric functions $\varepsilon_1(\omega)$ and $\varepsilon_2(\omega)$ for the materials sandwiching the graphene. These values are extracted by fitting of the hybrid plasmon-phonon dispersions (see the dotted curves in Figure 6b). It is worth mentioning that if we take the substrate dielectric function as a constant value,
independent of wavelength, we recover the $\sqrt{q}$ plasmon dispersion without plasmon-phonon hybridization (see the solid line in Figure 6b). Figure 6b also presents an intensity plot of the full-wave simulations based on the dielectric function with phonon features, which is overlaid on the resonance peak frequencies obtained from the experimental spectra. We find a good agreement from a comparison of the measured and calculated hybrid plasmon-phonon dispersions. It should be noted that for small $q$ ($\leq 0.5\times 10^5$\,cm$^{-1}$), the coupled plasmon-phonon dispersion has negligible deviation from the simple $\sqrt{q}$ dispersion. This agrees with previous experimental observations for graphene microstructures in this size range~\cite{Ju:2011}. More importantly, the full-wave simulations based on the dielectric functions of graphene and the substrate can describe the overall physics in this system, such as the excitation and evolution of the three hybrid plasmon-phonon resonances, resulting in two branch-splittings (Figure 6b). Furthermore, other features like the spectral weight transferring from the low- to high-frequency plasmon branch and the Fano-like resonance shape can also be extracted from the numerical simulations (see Supplementary Information, Figure S3).

\begin{figure}[t!]
\begin{center}
\includegraphics[width=1\columnwidth]{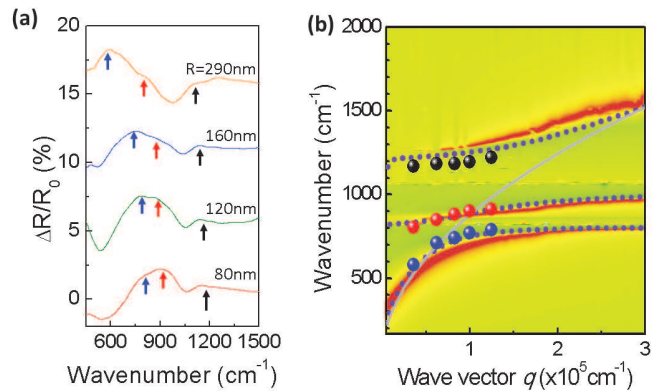}
\end{center}
\caption{Plasmon-phonon coupling and dispersion relations. ({\bf a}) Resonance spectra of graphene dot arrays with different dot radii on SiO$_2$/Si. The spectra are vertically displaced for clarity. ({\bf b}) Plasmon frequencies as a function of wave vector $q=1/R$ for peaks 1, 2 and 3 in ({\bf a}) and Figure 4b are plotted as dots. The theoretical calculation and numerical simulation of the plasmon-phonon hybridization are plotted as dotted branches and a two-dimensional pseudocolour background, respectively. The solid line represents the calculated plasmon frequencies without considering the effect of the substrate phonons. See Supplementary Information for details of our calculations and simulations.\label{fig6}}
\end{figure}

We have achieved large-area fabrication of nanoscale graphene dot and antidot arrays, which support plasmon resonances in the mid-infrared regime. Our approach is highly feasible for production of large-area graphene dot and antidot arrays with nanoscale features via a simple fabrication process. The extraordinary light confinement in the achieved plasmonic arrays has great potential for molecular sensing~\cite{Johnson:2010} based on enhanced infrared light interaction, for infrared light harvesting~\cite{Li:2011} as well as for spectral photodetection~\cite{Chitara:2011}.

Our study reveals the importance of polar substrates and their intrinsic optical phonons for both understanding and exploring the plasmon dispersion. Indeed, coupling between the substrate phonons and graphene plasmons is significant and may extend the optical tunability of graphene plasmons beyond that offered by electrostatic gating. The hybridized plasmon-phonon mode in the vicinity of the surface polar phonon frequencies is experimentally demonstrated and further validated by theoretical calculations and numerical simulations. These efforts have important implications for the understanding of graphene plasmonics and our findings hold potential for applications of graphene in high-performance mid-infrared photonic devices.

\emph{Acknowledgments.} This work was partly supported by the Catalysis for Sustainable Energy Initiative Center (CASE), funded by the Danish Ministry of Science, Technology and Innovation. The Center for Nanostructured Graphene (CNG) is sponsored by the Danish National Research Foundation, Project DNRF58.

\end{document}